\documentclass[aps,prc,reprint,showpacs,showkeys,amsmath,amssymb,longbibliography,lengthcheck,floatfix,nofootinbib]{revtex4-2}
\usepackage[dvips]{graphicx} 
\usepackage{dcolumn} 
\usepackage{amsmath}
\usepackage{pdfscreen} 
\begin{document} 
\setcounter{topnumber}{1}
\textfloatsep=0.1in
\renewcommand{\topfraction}{0.99}
\renewcommand{\bottomfraction}{0.99} 
\renewcommand{\textfraction}{0.01}
\title{Effect of potential-energy-model inaccuracies on predictions of
  fission-fragment mass distributions
  based on the  Brownian shape-motion method}
\author{Peter M\"{o}ller$^{1}$}\email{mollerinla@gmail.com}
\author{Christelle Schmitt$^{2}$}
 \affiliation{$^{1}$Mathematical Physics, Lund University, S-221 00 Lund, Sweden \\
$^2$Institut Pluridisciplinaire Hubert Curien (IPHC), Rue du Loess,
     67000 Strasbourg, France}
%
\begin{abstract}
  \begin{description}

\item[Background]
  Most actinide
  nuclides fission asymmetrically. A common explanation was that
  this division benefited from leading to fragments
  in the vicinity of the doubly
  magic $^{132}$Sn. It was tacitly assumed
  that lighter nuclides would all fission symmetrically
  because a similar situation could not occur there.
  However, a weakly asymmetric mass distribution was found
  for some pre-actinides at energies about 10 MeV above the fission
  saddle point by Itkis {\it et al.\/} [Yad. Fiz. {\bf 52} (1990) 944;
  Sov. J. Nucl. Phys.  53 (1991) 757; Nucl. Phys. {\bf 640} (1998) 375].
  A more recent experiment by Andreyev {\it et al.\/} performed in June 2008
  [Phys.\ Rev.\ Lett.\ {\bf 105}, 252502 (2010)] showed
  a strongly asymmetric mass distribution in fission of $^{180}$Hg
  at low excitation energies. This gave rise to a
new focus on fission properties in the ``below Pb'' region.  
M{\"{o}}ller and Randrup presented a comprehensive
calculation, based on the Brownian shape motion (BSM) method, of
fission-fragment charge distributions [Phys. Rev. C
  {\bf 91} (2015) 044316 ] which obtained  that ``a new region of
asymmetry'' appeared for approximately $95 \le N \le 115$ and
$ 75 \le Z \le 94 $. Available experimental results at the time,
except for the observation of symmetric fission of
$^{187}$Ir Itkis {\it et al.\/} [Yad. Fiz. {\bf 52} (1990) 944], 
agreed
with these predictions apart for minor differences in the transition
regions between predicted symmetric and asymmetric fission.
It was argued   [Phys. Rev. C
  {\bf 91} (2015) 044316 ] that the inaccurate results  for  $^{187}$Ir
were related to inaccuracies in the calculated potential-energy
surface and that such inaccuracies are related to the (in)accuracies
of the calculated ground-state masses for the corresponding mass splits.

\item[Purpose]
  We expand on our previous discussion
  of a possible source of the difference between the experimental
  and theoretical fission-fragment mass distributions for $^{187}$Ir
  and furthermore investigate if 
  such differences may occur for other fissioning nuclides
  in the ``below Pb'' and actinide regions.

\item[Methods]
  It has been shown that
  all structure in the mass distributions obtained by use of the BSM method
  is entirely due to the structure of the potential-energy surfaces
  on which the random walks are executed. Therefore, to understand the
  discrepancy between the previous theoretical results
  and the experimental observations for $^{187}$Ir we focus on the accuracy
  of the calculated potential-energy surface. 

\item[Results]
  We find that in symmetric fission of $^{187}$Ir the corresponding
  calculated fragment ground-state masses are too high compared to experiment
  by about 2.5 MeV each so the scission potential energy is calculated to
  be too high by  5 MeV\@. We also find that this is  the
  largest error that can occur for any scission configuration in heavy-element
  fission.  

\item[Conclusions]
As earlier we  pose that the reason that symmetric fission is not favored in the
calculated fission-fragment distribution of $^{187}$Ir is that the
potential-energy surface is overestimated by 5 MeV for symmetric splits.
In the calculations a large error for symmetric splits
extends only to nuclides a few nucleons beyond
$^{187}$Ir and does not occur elsewhere.
Therefore, to test this hypothesis of the origin
of the discrepancy, it is of interest to map out in
experiments how far this
region of symmetric fission ``within the predicted region of
asymmetry'' extends and if substantial discrepancies occur elsewhere.
   \end{description} 
 \end{abstract}
\maketitle
\mbox{ } \\[-0.01in]
\section{INTRODUCTION}
Fission below  the actinide region has historically not been studied
as extensively as fission of the heaviest nuclei. It was tacitly assumed
that because of the absence of strong shell corrections in possible fragment
splits of these nuclei they would all obey ``liquid-drop'' systematics,
fission symmetrically, and therefore be fairly uninteresting for
studying details of the fission process.
\begin{figure*}[t] 
 \begin{center} 
 \includegraphics[width=0.95\linewidth]{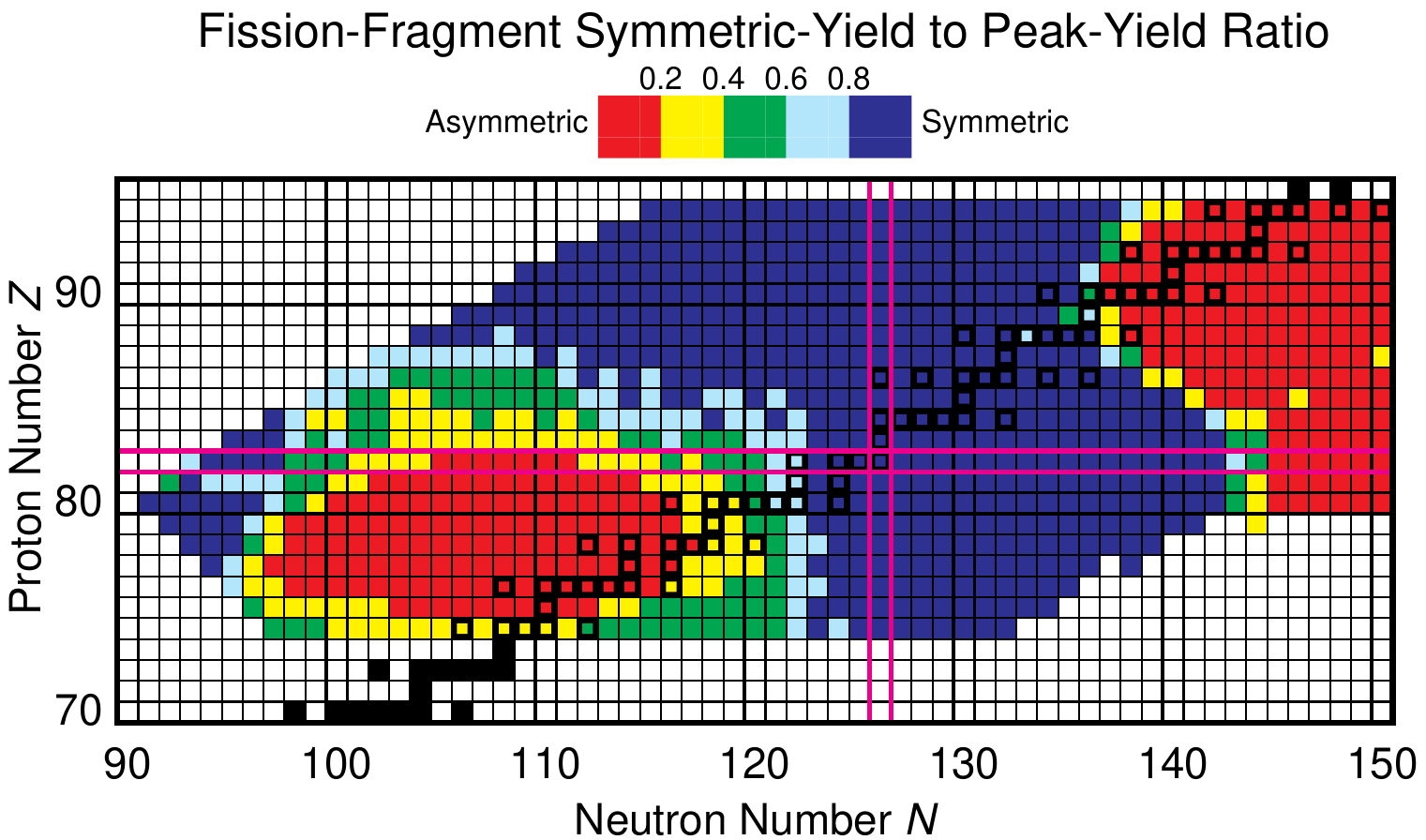} 
 \caption{(Color online) Calculated
   symmetric-yield to peak-yield ratios for 987
fissioning systems. Black open squares 
indicate $\beta$-stable nuclei.
These previously obtained results \protect\cite{moller15:b}
show a new, contiguous region of asymmetric fission separated
from the classical location of asymmetric fission in the actinides  by
an extended area of symmetric fission. The location of $^{187}$Ir is
 well inside the red region.}  
\label{chartasym} 
 \end{center} 
\end{figure*} 	
\mbox{ }
The discovery, in June of 2008, that fission of $^{180}$Hg at the  particularly low
excitation  energies
($E^* < 10$ MeV) populated in $\beta$-decay of $^{180}$Tl
results in a well-developed
asymmetric fragment mass distribution \cite{andreyev10:a}
put an abrupt end to this expectation.

Subsequently, experimental studies started to focus
on nuclei in the ``sub-Pb'' (and slightly above) region; some
early studies are
those of Refs.\  \cite{liberati13:a,elseviers13:a}. 
About two years after the asymmetric fission of $^{180}$Hg was observed,
Randrup {\it et al.\/} \cite{randrup11:a} developed the 
BSM method, which allowed fission mass distributions to be calculated
(for excitation energies above the barrier) for any
fissioning compound nucleus. It was
based on random walks on previously calculated
five-dimensional potential-energy
surfaces calculated as functions of five shape
parameters \cite{moller01:a,moller09:a}.
\begin{figure}[b] 
 \begin{center} 
 \includegraphics[width=0.95\linewidth]{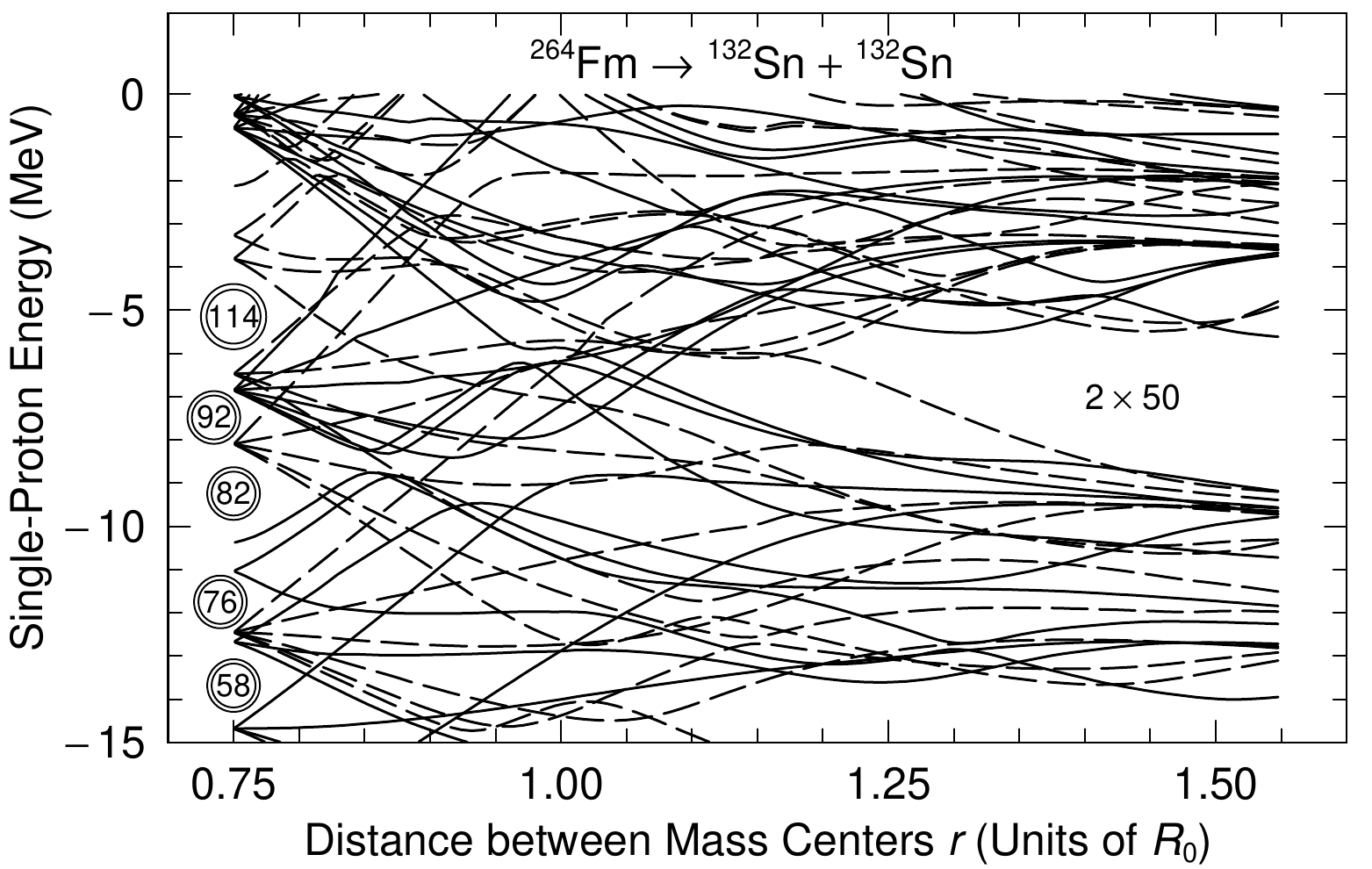} 
 \caption{Calculated single-particle levels for shape evolution
   from spherical
   shape to separated symmetric spherical fragments.}  
\label{264fmplev} 
 \end{center} 
\end{figure}

The Brownian shape-motion model
\cite{randrup11:a,randrup11:b,randrup13:a}  in
its initial implementation \cite{randrup11:a},  had no {\it adjustable}
parameters.  There are two parameters, the strength of the bias
potential and the critical neck radius at which we assume the
fission-fragment mass asymmetry is frozen in. Since it has been  shown that
the results are insensitive to a large range of these two parameters
\cite{randrup11:b}, they are not in the category of
adjustable parameters.  In the version of the model used subsequently,
in particular in the large-scale calculation showing the extension
of the ``new region of asymmetry'' \cite{moller15:b}
there
are two adjustable parameters that  govern the rate
at which the shell effects dampen out with energy determined in Ref.\
\cite{randrup13:a}.  The model has been 
extensively benchmarked, in Ref.~\cite{randrup13:a} with respect
to 70 charge yields measured at GSI \cite{schmidt00:a}, and in
Ref.~\cite{ghys14:a} with respect to new and older data in the neutron-deficient
Pb region.
It is worth noting that the model has been extended and applied to
additional fission properties such as neutron emission correlation with
fission fragment masses \cite{albertsson21:a}, odd-even staggering,
and fission-fragment isotopic
yields \cite{moller15:c,moller17:a,schmitt21:a}.
\begin{figure*}[t] 
 \begin{center} 
 \includegraphics[width=0.95\linewidth]{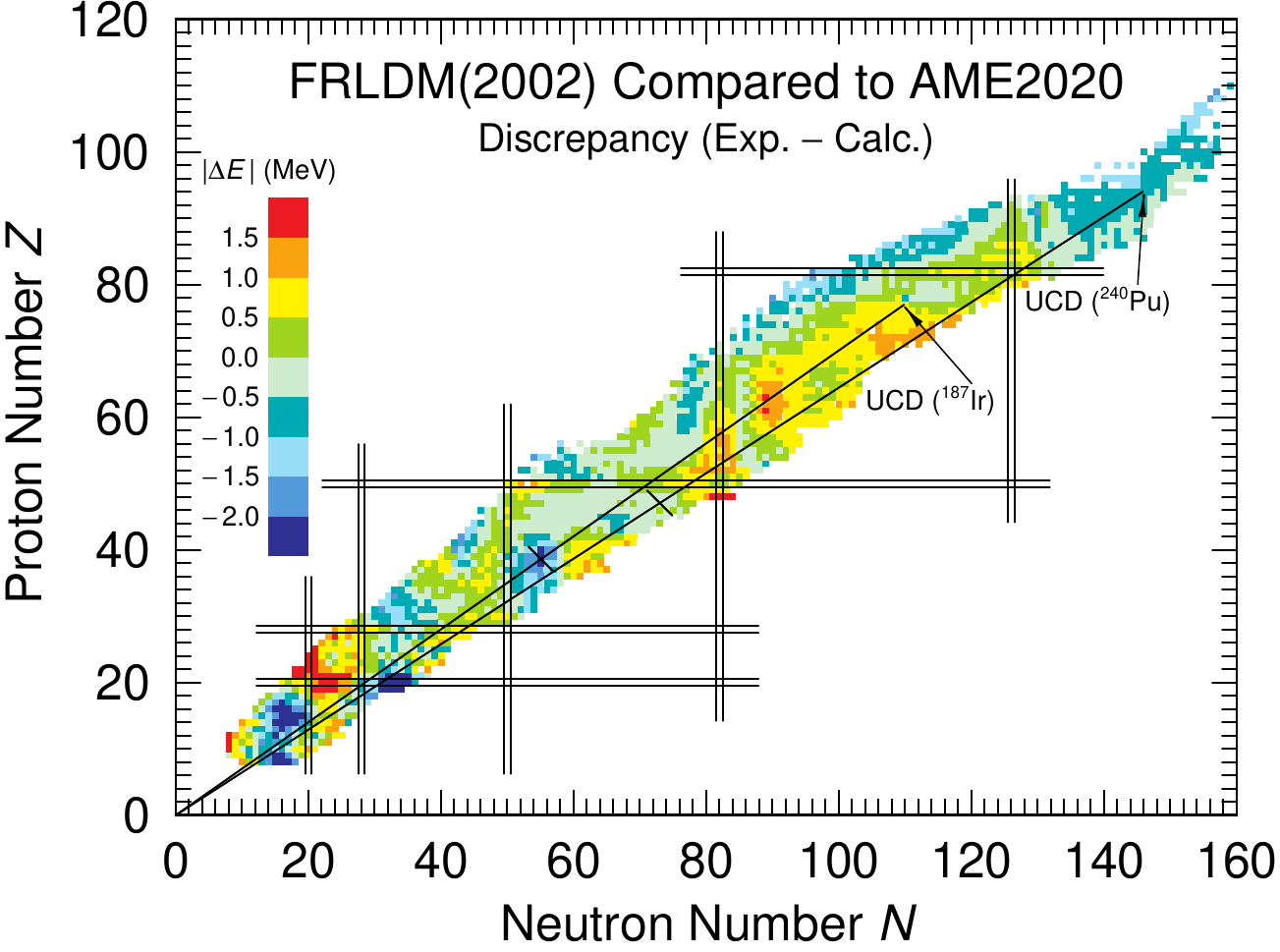} 
 \caption{(Color online) Difference between experimental masses 
   and theoretical  masses
   calculated in the FRLDM2002 \protect\cite{moller04:a}, which was
   the potential-energy model used to calculate the 5D potential-energy
   surfaces (it was the most current potential-energy model in the time-frame  2004--2006). Superimposed
   are the UCD lines for fission of $^{187}$Ir and $^{240}$Pu with symmetric
 fragment locations indicated by  short black lines.}
\label{devwithucd} 
 \end{center} 
\end{figure*} 	
\mbox{ }

\begin{figure}[t] 
 \begin{center} 
   \includegraphics[width=0.95\linewidth]{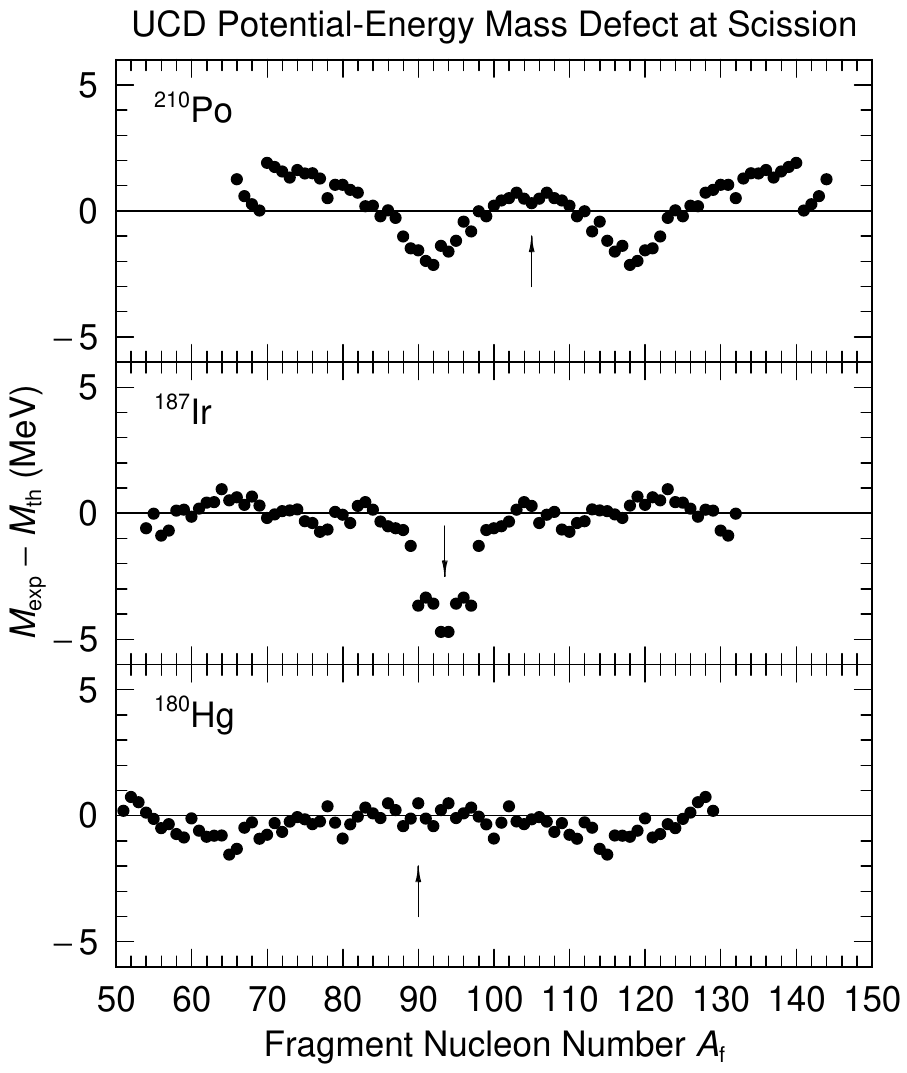} 
 \caption{Difference between experimental masses and calculated ground-state
   masses for a range of mass splits in fission along the UCD lines for three
   representative compound systems
   in the ``below Pb region''.} 
\label{dev3nuclow} 
 \end{center} 
\end{figure} 	
\mbox{ }

\begin{figure}[t] 
 \begin{center} 
 \includegraphics[width=0.95\linewidth]{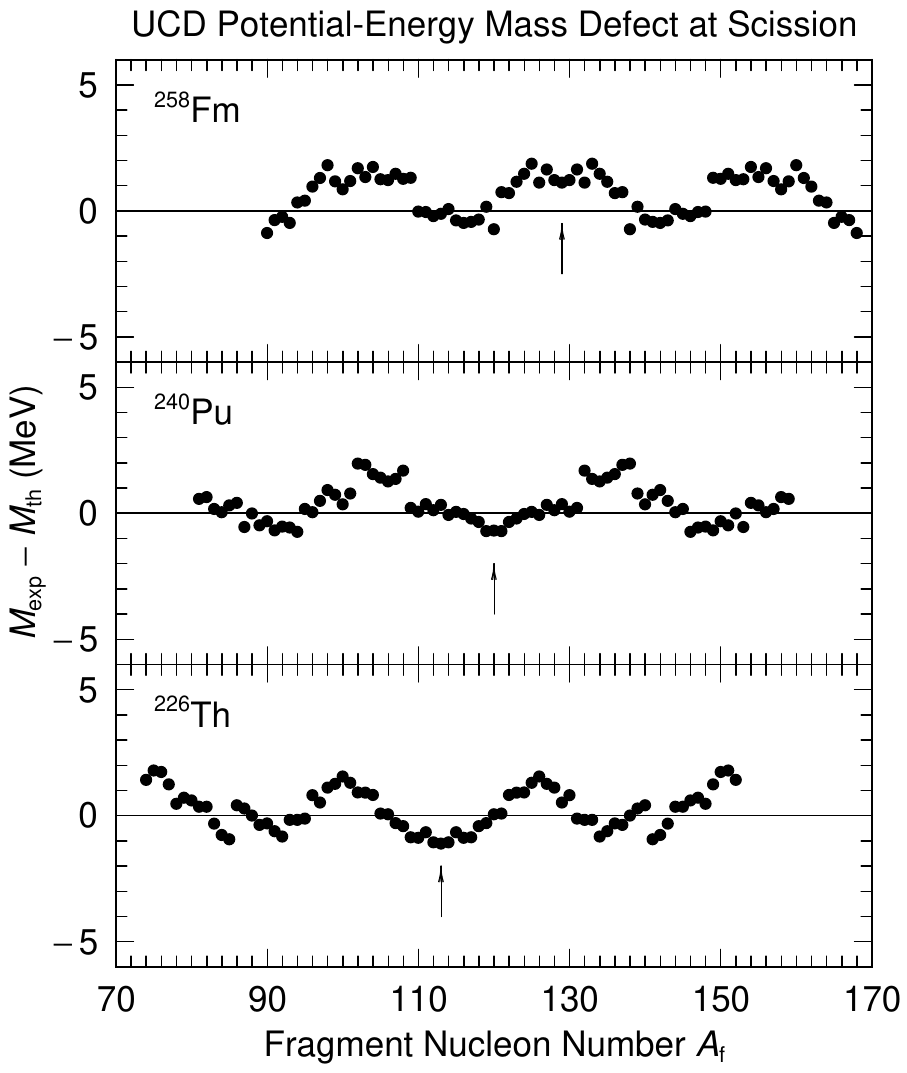} 
   \caption{Difference between experimental masses and calculated ground-state
   masses for a range of mass splits in fission along the UCD lines for three
   representative compound systems in the ``classical region of asymmetry''
   (actinide region) .}  
\label{dev3nucact} 
 \end{center} 
\end{figure} 	
\mbox{ }

\begin{figure*}[t] 
 \begin{center} 
 \includegraphics[width=0.95\linewidth]{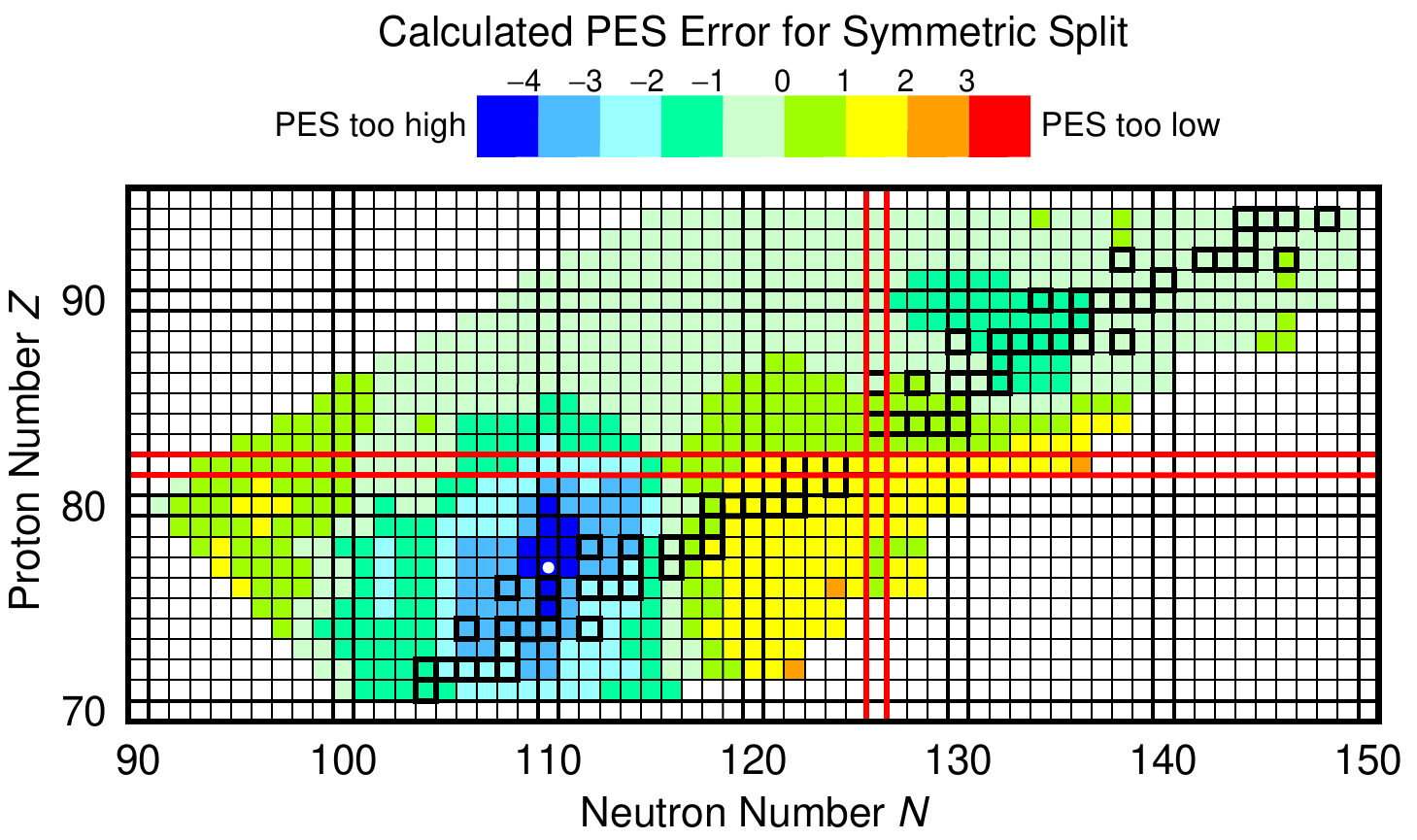} 
 \caption{(Color online)
Difference between experimental masses and calculated ground-state
masses for symmetric mass splits for fission of compound systems in
the ``below Pb region'' and extending
into the beginning of the actinide region. The location of the compound fissioning
nuclide $^{187}$Ir is indicated by a
white circle. Open black squares indicate experimentally
known $\beta$-stable nuclides.}  
\label{devsymfisslow} 
 \end{center} 
\end{figure*} 	
\mbox{ }

\begin{figure*}[t] 
 \begin{center} 
 \includegraphics[width=0.95\linewidth]{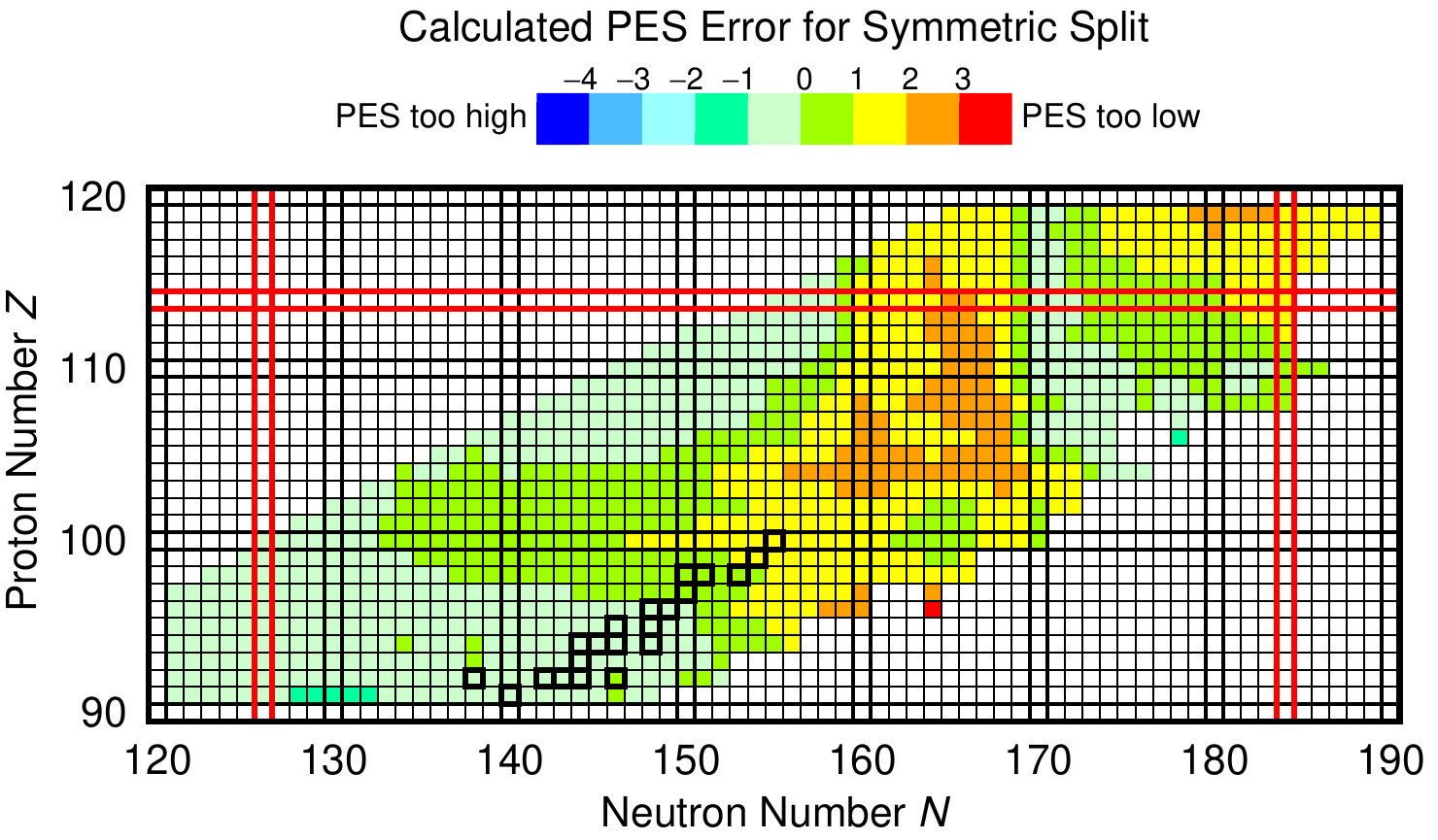} 
 \caption{(Color online)Difference between experimental masses and calculated ground-state
masses for symmetric mass splits for fission of compound systems in the region
of the heaviest elements.  Open black squares indicate
experimentally known $\beta$-stable nuclides.
}  
\label{devsymfisshigh} 
 \end{center} 
\end{figure*} 	
\mbox{ }

There are few theoretical approaches to calculate
fission-fragment mass distributions and other fission properties,
that have so far been routinely applied
to basically all fissioning systems and which show overall good quantitative  agreement
with observations. In contrast, after the initial sensitivity studies
and benchmarking of the BSM method, it was applied to the calculation of
fission-fragment mass yields for 987 nuclides in the region $ 74 \le Z \le 94$,
$91\le N \le 150$. Adjacent to $^{180}$Hg the calculations showed a new
contiguous region of asymmetric fission
for approximately  $95 \le N \le 115$ and
$ 75 \le Z \le 94 $ \cite{moller15:b}. In the paper it was noted that
the observation by Itkis and collaborators \cite{itkis90:a,mulgin98:a}
of the symmetric fission of $^{187}$Ir contradicted the calculated results.
It was proposed that the discrepancy was due to inaccuracies in
the calculated 5D potential-energy surface for $^{187}$Ir for trajectories leading to symmetric division
and that the magnitude of the inaccuracies of the calculated potential-energy
surface could be related to the inaccuracies of the calculated  ground-state
masses of the corresponding fission fragments. The recent experimental paper by Duhri {\it et al.\/}  \cite{dhuri22:a} revisited in more detail the fission 
properties of $^{187}$Ir and commented on the discrepancy with respect to
the studies in \cite{moller15:b}, which motivates us to
present more detailed arguments for the proposed origin of the inaccuracy
of the results for $^{187}$Ir and if this can occur for other fissioning nuclides.

\section{Regions in the nuclear chart where  the BSM method may be
inaccurate}
The BSM method is a random walk on 5D potential-energy surfaces previously
calculated
as functions of five nuclear shape ``coordinates'', namely elongation,
neck radius, left and right fragment ellipsoidal deformations,
and mass (or equivalently volume) left-right fragment asymmetries.
As implemented in the calculations shown in Fig.\ \ref{chartasym},
all structures present in the fission-fragment mass distributions 
are consequences of the structures in the calculated potential-energy
surfaces. Therefore, to shed light on the possible origin of
the different experimental and theoretical results for the fission-fragment
mass distributions  for $^{187}$Ir and if they can occur elsewhere in
the nuclear chart we need to understand
the accuracy of the calculated
potential-energy surfaces. However, the potential-energy surfaces consist of
more than 5 million values corresponding to different shapes.
Experimental values to compare to do not exist for these points except
for the ground state, and to a lesser extent for the calculated
fission-barrier saddle points and isomeric minima.

\subsection{Relation between 5D potential-energy surface accuracy and
  fission-fragment ground-state masses}

At scission when the fission fragments have almost completely
formed it is reasonable to assume that at this point
the 5D potential-energy surface
accuracy is  closely related to the accuracy of the model as applied to the
ground-state masses of the fragments involved. We furthermore know that in our model
the level structure at scission appears long before
the scission point, obviously where the nascent fragment shape is
similar to the shapes at scission. This is seen in Fig.\ \ref{264fmplev}
(Fig.\ 7 in Ref.\ \cite{moller87:c}),
where the fragment $Z=50$ proton magic-number gap has almost fully developed
already at
$r=1.25$ which corresponds to the value at the fission isomeric shape;
see Ref.\ 
\cite{moller87:c} in which the  potential-energy surfaces in Fig.\ 6  show this.
Therefore it is reasonable to assume that the potential-energy surface
already long before scission has inaccuracies whose magnitude increases with
increasing inaccuracies in the calculated masses of the corresponding fragments.
These observations  allow us to form an opinion where in the chart
the calculated fission-fragment mass distributions are less accurate and in
possible disagreement with experiment,  as is the case for $^{187}$Ir.

\subsection{Estimates of BSM method inaccuracies  across the heavy-element
  region}
We have called the mass model based on the parameters used in the
potential-energy calculation FRLDM2002. It was not published,
because it was an ``interim'' mass table,
but the most current and best  we had at the
time the potential-energy calculations were carried out.
For more details  see
Ref.\ \cite{moller04:a}.
For the most
current mass tables  see
Ref.\ \cite{moller16:a}. We compare in Fig.\ \ref{devwithucd}
FRLDM2002 to experimental masses. In the figure we also show two UCD
lines, one for fission of $^{240}$Pu and the other for $^{187}$Ir. UCD
stands for unchanged charge distribution, that is the $N/Z$ ratio in the fragments is
the same as in the fissioning system. This assumption is used in the
standard version of the BSM method in
order to partition the neutrons and protons
between the two fragments for a given mass split.   The location
of the symmetric split is indicated by short lines crossing the UCD
lines.  When we discuss errors along the UCD line and when  this line
does not locate on integer proton or neutron numbers we take the average error
of the most nearby masses (with integer $Z$ and $N$).

For fission of $^{187}$Ir symmetric fission leads to fragments for
which the masses are calculated to be too high by about 2.5 MeV. Since
the symmetric fission configuration consists of two such fragments we
find that the potential-energy surface is too high by about 5 MeV for
the symmetric fission configuration for $^{187}$Ir.
In contrast, for
symmetric fission of $^{240}$Pu the mass discrepancy at symmetry is
almost zero. We illustrate in Figs.\ \ref{dev3nuclow}--\ref{dev3nucact}
the potential-energy accuracy
determined in this way along the entire UCD lines of six
representative fissioning systems, namely
$^{180}$Hg, $^{187}$Ir, $^{210}$Po, $^{226}$Th, $^{240}$Pu, and $^{258}$Fm.
This type of plot shows if a particular
asymmetric
mass split might be favored (positive deviation) or disfavored compared to
a more accurate potential-energy surface.
At each fragment split $A_{\rm f}$ the plotted error is the sum of the calculated mass
error of the two fragments.
We note immediately that
the largest deviation is for symmetric split of $^{187}$Ir,
namely $\approx -5$ MeV\@. With our definition of potential-energy surface
accuracy at scission this is actually the largest deviation that can
occur for symmetric scission points for any nuclide in the region
$Z \ge 70$ and $N \ge 90$ because there are no larger mass errors in
the FRLDM2002 mass model in
the regions where the fission fragments ``land''.
Because at each division  we have added the error of the light
and heavy fragment the sum can be small when the individual fragment
errors are large when they are of opposite sign and roughly cancel each other.

We show in Figs.\ \ref{devsymfisslow} -- \ref{devsymfisshigh}
the potential-energy accuracy at symmetric scission configurations
for all heavy nuclei when experimental
masses for the fragments are available.
As surmised just above, the largest deviations occur
in the vicinity of the location of the symmetric split of $^{187}$Ir.
We also notice that in the heavier region from the actinide to the superheavy
region the errors in the landing locations for symmetric fission
are consistently much smaller. This is also to be expected because
fission fragments in symmetric fission  of these heavy nuclides land in
a heavier mass region where the mass-model error is considerably smaller than in the
landing region in fission of rare-earth nuclides. 
As a consequence, no major difference between calculated results and experimental observations
were observed for fission of actinides,
as benchmarked in \cite{randrup11:a,randrup13:a}. However some larger deviations in
detailed, calculated fission-fragment charge distributions that include descriptions
of odd-even staggering are visible in results in Ref.\ \cite{moller17:a}, in particular
near $Z=36$ in the light fragment peak in fission of uranium isotopes.  It
was pointed out that this deviation also occurs because 
the corresponding light fragments land in the region of large mass-model deviations centered at $Z=40$ and $N=56$.

\section{Summary and Conclusions}

The actinide region of asymmetric fission  was extensively studied for  close to
a century. In contrast, the extent of the
pre-actinide asymmetric-fission "island" is under very active research
only since the time of the experiment of Andreyev {\it et al.\/} in 2008 \cite{andreyev10:a},
which demonstrated the
occurrence of a well-developed asymmetric split for fission at low excitation energies
of the
neutron-deficient nuclide $^{180}$Hg.
In 2015  the BSM method, which
gave realistic results for actinide fission, in a large-scale calculation
predicted a new contiguous region of asymmetric fission in the
``rare-earth''  (also referred to as ``sub-Pb'') region.
These predictions were found to be overall consistent with the numerous
experimental studies going on in parallel since the communication by
Andreyev {\it et al.\/} \cite{andreyev10:a}. However, recently,
from the experimental side, Dhuri {\it et al.\/} \cite{dhuri22:a} emphasized the clear
discrepancy between the model and  measurements for $^{187}$Ir. The
calculation  showed a well-developed asymmetric fission-fragment mass distribution at low excitation energy,
whereas experiments are dominantly symmetric.

Symmetric fission fragments in the fission of $^{187}$Ir land in an area
where the mass-model errors and consequently the potential-energy errors
near scission are unusually large. It is well-known since more than
half a century that in nuclei with proton number near $Z=40$ and  $N=56$
the corresponding spherical level gaps open up. These two gaps become much larger
than has so far been obtained in global single-particle models.
In nuclei  symmetric fission leading to
fragments near these nucleon numbers is therefore unusually
favored relative  to  theory. Errors the calculated 5D potential-energy surfaces
leading to fragment splits away from this region for any other fissioning nuclei
are probably less than half of this error, that is less than half the
5 MeV we see in the potential-energy
surface corresponding to symmetric division of $^{187}$Ir.
Only a few fissioning systems can have {\it both}
fragments land in this region of large errors (and it must obviously
be symmetric fission) which are seen in Fig.\ \ref{devwithucd}.
In Figs.\ \ref{dev3nuclow} and \ref{dev3nucact} there are deviations of
up to $\pm 2$ MeV for $^{180}$Hg, $^{210}$Po, $^{226}$Th, $^{240}$Pu, and $^{258}$Fm.
However BSM method calculations of fission-fragment mass distributions
for these nuclides agree quite well with experiment
\cite{randrup11:a,randrup13:a,albertsson20:a}.
Therefore local inaccuracies up to this magnitude do not seem to
make the calculated fragment distributions inaccurate. Our discussions in
this paper showed that larger calculated 5D potential-energy surface inaccuracies
occur only in the vicinity of $^{187}$Ir. Therefore we expect that calculated
fragment mass  distributions will agree with (future) experiments elsewhere.

We are grateful to Kripamay Mahata for discussions about the 
$^{187}$Ir results and comments on this manuscript.


\end{document}